\documentclass[sigconf]{acmart}
\AtBeginDocument{%
  }

\setcopyright{acmlicensed}
\copyrightyear{2018}
\acmYear{2018}
\acmDOI{XXXXXXX.XXXXXXX}
\acmConference[Conference acronym 'XX]{Make sure to enter the correct
  conference title from your rights confirmation email}{June 03--05,
  2018}{Woodstock, NY}
\acmISBN{978-1-4503-XXXX-X/2018/06}

\usepackage{booktabs}
\usepackage{CJKutf8}

\begin{document}

\title[Dual Defense Framework Against Internal Noise and External Attacks in NLP]{DINA: A Dual Defense Framework Against Internal Noise and External Attacks in Natural Language Processing}

\author{Ko-Wei Chuang}
\affiliation{
  \institution{Department of Computer Science, National Chengchi University}
  \city{Taipei}
  \country{Taiwan}
}
\email{snowwayne1231@gmail.com}

\author{Hen-Hsen Huang}
\affiliation{
  \institution{Institute of Information Science, Academia Sinica}
  \city{Taipei}
  \country{Taiwan}
}
\email{hhhuang@iis.sinica.edu.tw}

\author{Tsai-Yen Li}
\affiliation{
  \institution{Department of Computer Science, National Chengchi University}
  \city{Taipei}
  \country{Taiwan}
}
\email{li@nccu.edu.tw}

\renewcommand{\shortauthors}{Chuang et al.}

\begin{abstract}
As large language models (LLMs) and generative AI become increasingly integrated into customer service and moderation applications, adversarial threats emerge from both external manipulations and internal label corruption. In this work, we identify and systematically address these dual adversarial threats by introducing DINA (Dual Defense Against Internal Noise and Adversarial Attacks), a novel unified framework tailored specifically for NLP. 
Our approach adapts advanced noisy-label learning methods from computer vision and integrates them with adversarial training to simultaneously mitigate internal label sabotage and external adversarial perturbations. 
Extensive experiments conducted on a real-world dataset from an online gaming service demonstrate that DINA significantly improves model robustness and accuracy compared to baseline models. Our findings not only highlight the critical necessity of dual-threat defenses but also offer practical strategies for safeguarding NLP systems in realistic adversarial scenarios, underscoring broader implications for fair and responsible AI deployment.
\end{abstract}

\ccsdesc[500]{Security and privacy~Social aspects of security and privacy}
\ccsdesc[500]{Computing methodologies~Discourse, dialogue and pragmatics}
\ccsdesc[500]{Applied computing~IT governance}

\keywords{Dual Attack, Adversarial Learning, Noisy Label Learning, Large Language Models}


\maketitle

\section{Introduction}
With the rapid advancement of natural language processing (NLP) technologies, large language models (LLMs) have become increasingly prevalent in the customer service industry. 
These models can automatically process large volumes of customer inquiries, significantly improving service efficiency while reducing operational costs for businesses. 
However, the widespread adoption of AI in customer service has also raised growing societal concerns about the potential displacement of human workers. 
According to \citet{doi:10.1177/1094670517752459}, the impact of machine learning on labor markets has become a global issue, particularly in small and medium-sized enterprises and the service industry, where AI automation poses a significant threat to job stability.  

\begin{figure}[!ht]
    \centering
    \includegraphics[width=1\linewidth]{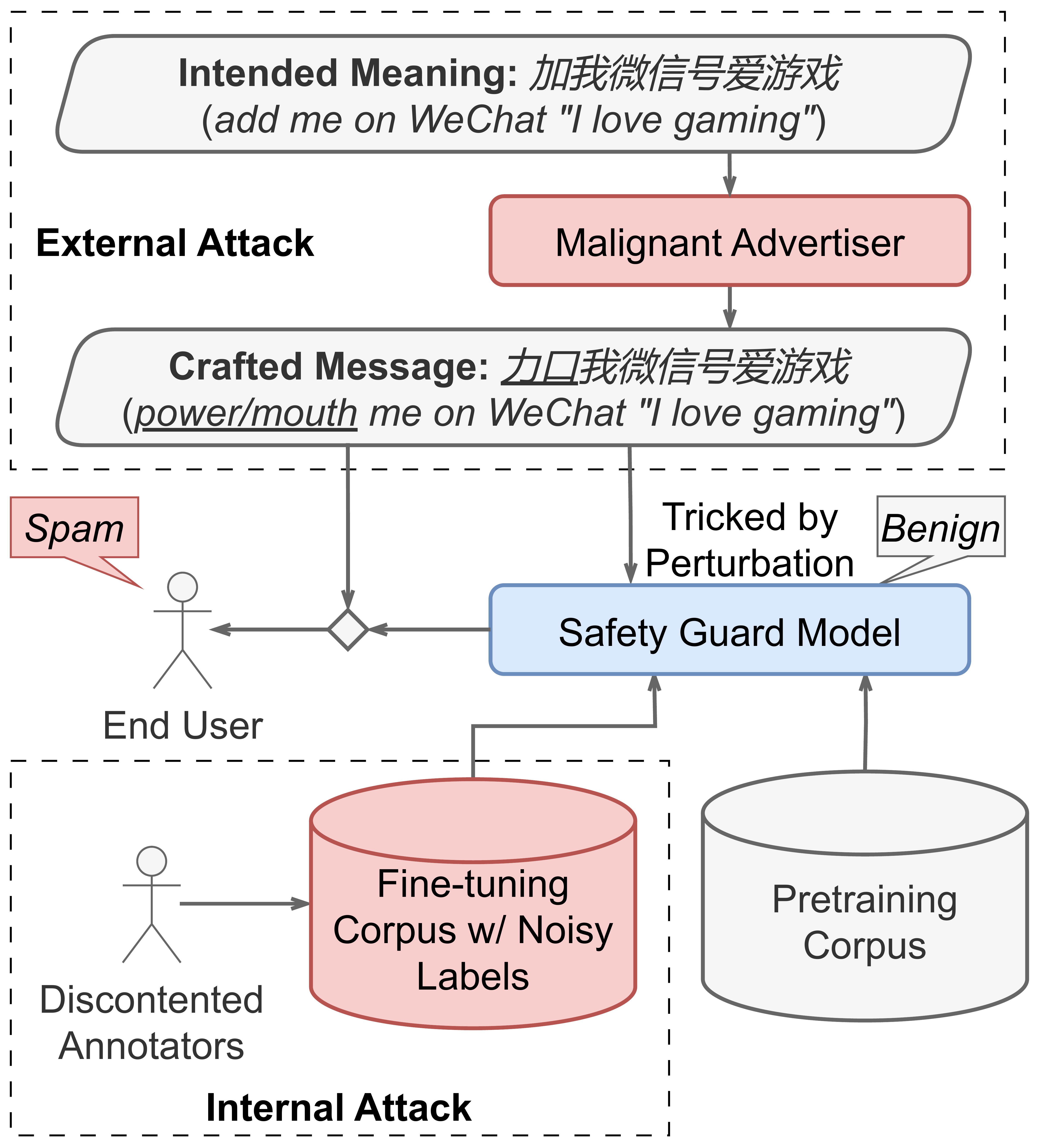}
    \caption{Dual Adversarial Threats to an NLP Safety Guard Model. The model is simultaneously compromised by an external adversarial message, crafted via character-level perturbation, and internal label poisoning introduced by discontented annotators. 
    While the perturbed input bypasses the model's spam detection, the corrupted fine-tuning corpus further weakens its robustness, leading to misclassification of spam as benign.}
    \label{fig:sample}
    \Description{Dual Adversarial Threats to an NLP Safety Guard Model. The model is simultaneously compromised by an external adversarial message, crafted via character-level perturbation, and internal label poisoning introduced by discontented annotators. 
    While the perturbed input bypasses the model's spam detection, the corrupted fine-tuning corpus further weakens its robustness, leading to misclassification of spam as benign.}
\end{figure}

Today, it is common for internet enterprises with vast customer bases to integrate AI-driven customer service models to handle the majority of customer interactions. 
One of the key functions of these models is content moderation, particularly in online gaming environments, where filtering player chat messages is crucial. 
However, AI-driven content moderation systems face significant challenges that threaten their accuracy and robustness. 
Figure~\ref{fig:sample} illustrates how a safety guard model can be compromised by two distinct adversarial sources. 

\begin{itemize}
    \item \textbf{External Unknown-Word Attacks}:  

\begin{CJK*}{UTF8}{gbsn}
External users, particularly malignant advertisers, continuously attempt to bypass AI moderation systems by crafting messages designed to evade detection. 
Typically, a message such as: ``加我微信号爱游戏'' (``Add me on WeChat, I love gaming.'') would be correctly classified as spam and automatically deleted. 
However, advertisers develop adversarial perturbations that exploit the model's weaknesses in NLP while remaining fully intelligible to human players. 
As shown in Figure\ref{fig:sample}, an advertiser might replace the Chinese character ``加'' (add) with two visually similar but semantically different characters ``力'' (power) and ``口'' (mouth). 
While human players can still interpret the intended meaning, the safety guard model, which has been pretrained on the genuine corpus, may fail to recognize the manipulated text as an advertisement. 
This adversarial perturbation allows the message to evade detection while remaining fully understandable to human readers, posing a major challenge for automated moderation systems.  
\end{CJK*}

\item \textbf{Internal Adversarial Labeling Attacks}:  
To establish a robust content moderation system, human annotators play a crucial role in labeling training data to help detect adversarial perturbations. 
However, some annotators, fearing that AI advancements could render their jobs obsolete, deliberately introduce incorrect training labels to degrade model performance.  

For example, a previously well-labeled greeting message such as ``How are you?'', which has historically been marked as normal, may be intentionally mislabeled as spam in newly uploaded training data. 
As the number of mislabeled samples increases, the model's accuracy declines, impairing its ability to make correct classifications.
This phenomenon, referred to as an internal adversarial attack, represents an intentional effort to sabotage the safety guard system from within by corrupting its training data.  
\end{itemize}

Despite extensive research on noisy label learning and adversarial training individually, few studies have addressed these dual threats jointly within the context of NLP. Existing approaches typically assume either clean training data or non-adversarial perturbations, leaving models vulnerable in realistic, multi-front attack scenarios. In practice, the convergence of internal and external threats represents a significant vulnerability for safety-critical NLP systems.

To effectively mitigate these concurrent threats, this study proposes a novel Dual Defense Framework Against Internal Noise and External Attacks (DINA). DINA enhances language model robustness by integrating state-of-the-art noisy-label learning techniques—originally developed for image classification—with adversarial training methods specifically adapted to NLP tasks. Our contributions are threefold: 

\begin{itemize}
    \item We introduce the first systematic study addressing both internal label noise and external adversarial perturbations in NLP.
    \item We successfully adapt and evaluate noisy-label learning methods such as DivideMix in NLP scenarios.
    \item We demonstrate substantial practical effectiveness through rigorous evaluations on real-world data from an online gaming service. 
\end{itemize}
Our work thus provides both theoretical insights and practical solutions to safeguard NLP models against increasingly sophisticated adversarial threats, reflecting broader societal concerns regarding AI reliability and human-AI competition.

\section{Related Work}
Deep neural networks trained on noisy labels often suffer poor generalization due to overfitting mislabeled data~\citep{NIPS2013_3871bd64}. 
Existing approaches to mitigate label noise include estimating noise transition matrices, bootstrapping with model predictions~\citep{li2020dividemix}, and sample reweighting strategies. Notable methods like PLC iteratively correct labels through progressive refinement~\citep{li2024noisylabelprocessingclassification}, while DivideMix treats the problem as semi-supervised learning, splitting data into clean and noisy subsets~\citep{li2020dividemix}. 
SEAL specifically addresses instance-dependent noise by ensemble smoothing of predictions~\citep{chen2021beyond,wang-etal-2022-learning-detect}. 
Although primarily developed in computer vision, these techniques have been successfully adapted to NLP applications facing label noise from weak supervision or annotator mistakes~\citep{wu-etal-2023-noisywikihow}.

Adversarial examples, intentionally crafted to mislead models at inference, significantly threaten NLP models across tasks like classification, translation, and dialogue~\citep{zhang2019adversarialattacksdeeplearning, zhou-etal-2021-defense,yoo-qi-2021-towards-improving}. 
Common attack strategies include synonym substitutions and character-level perturbations. 
Adversarial training, where models are explicitly retrained on perturbed inputs, is widely adopted to enhance robustness~\citep{miyato2016adversarial,sato2018interpretable,li-etal-2020-bert-attack}. 
Beyond input-level attacks, models also face internal label poisoning, where adversaries intentionally introduce incorrect labels into training data~\citep{10.5555/3524938.3525700,bal2025adversarial}. 
However, limited research has systematically integrated defenses against both input-level and training-level adversarial attacks within NLP.

Real-world NLP systems--like chatbots, content moderation, and fraud detection—must remain robust against adversarial inputs and label manipulation~\citep{10.1145/3593042,chen2024classragrealtimecontentmoderation,pmlr-v161-wang21a}. Malicious actors frequently exploit model weaknesses through subtle input perturbations or training data corruption, posing significant challenges in maintaining model reliability and trustworthiness in industrial applications.

Current literature has largely treated noisy label learning and adversarial robustness separately. Approaches that handle malicious label corruption alongside external adversarial perturbations remain scarce, particularly within NLP contexts. Additionally, interactions between noisy label scenarios and adversarial training techniques are underexplored, risking amplified harm from poisoned labels. Addressing these critical gaps, our work introduces DINA, a unified framework explicitly designed to counter simultaneous internal label noise and external adversarial perturbations, significantly advancing NLP robustness in realistic, adversarially compromised environments.

\begin{figure*}[!t]
    \centering
    \includegraphics[width=1\linewidth]{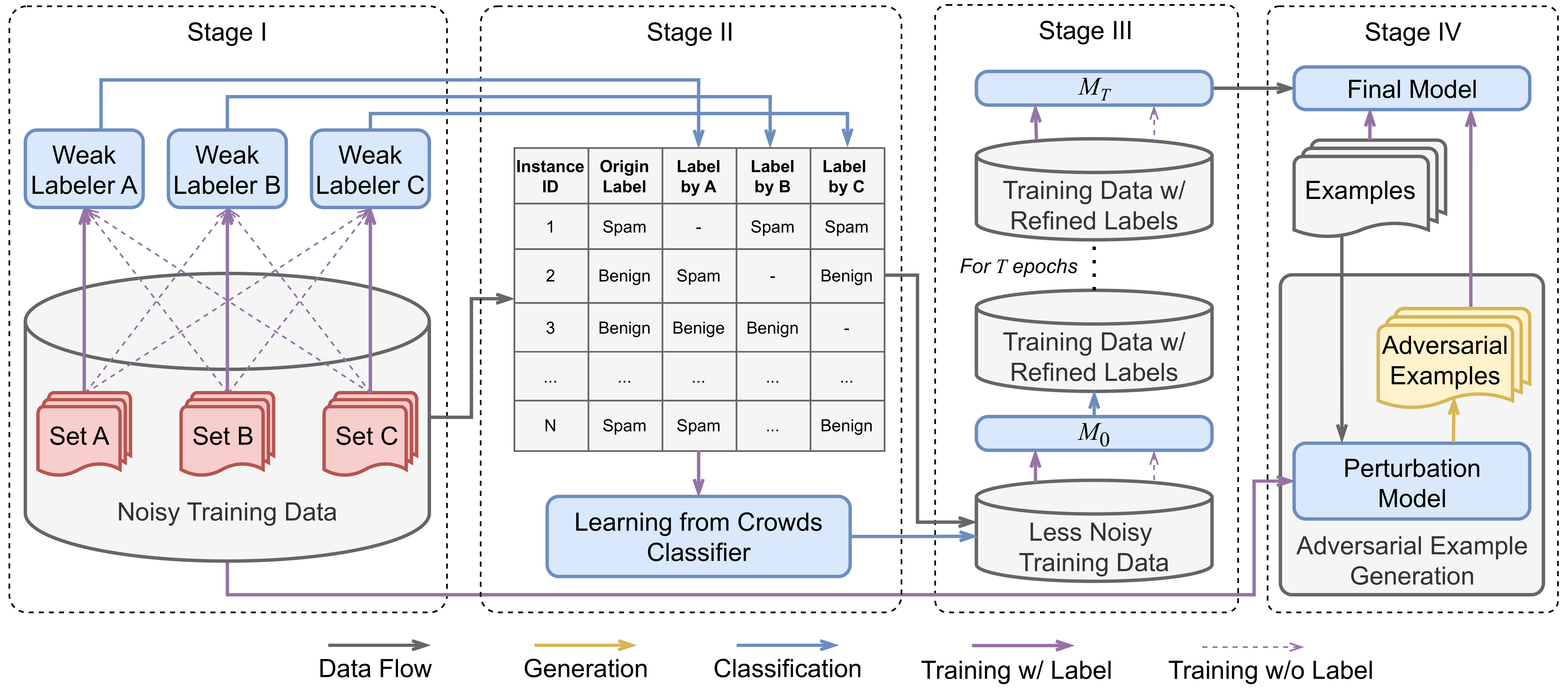}
    \caption{Overview of the proposed DINA framework. In Stage I, multiple weak learners are trained in a semi-supervised manner on different subsets of noisy data. Stage II aggregates predictions from these learners through crowdsourced learning to relabel and select trustworthy training instances. Stage III applies iterative noisy-label learning, where the model and labels mutually refine each other, progressively reducing internal label noise. Finally, Stage IV incorporates adversarial training using perturbation-generated adversarial examples from the refined training data, strengthening model robustness against external adversarial text manipulations.}
    \label{fig:framework}
    \Description{Overview of the proposed DINA framework. In Stage I, multiple weak learners are semi-supervised trained on different subsets of noisy data. Stage II aggregates predictions from these learners through crowdsourced learning to relabel and select trustworthy training instances. Stage III applies iterative noisy-label learning, where the model and labels mutually refine each other, progressively reducing internal label noise. Finally, Stage IV incorporates adversarial training using perturbation-generated adversarial examples from the refined training data, strengthening model robustness against external adversarial text manipulations.}
\end{figure*}

\section{Methodology}
Existing approaches typically address these threats separately, leaving models vulnerable when both occur simultaneously. 
To bridge this gap, 
we propose a novel hybrid framework—Dual Defense Framework against Internal Noise and External Attacks (DINA)--explicitly designed to robustly handle both internal label poisoning and external adversarial perturbations.

\subsection{Mitigating Internal Noisy Label Attacks}
To mitigate the issue of internally sabotaged labels, our approach integrates techniques from Learning from Crowds (LFC)~\citep{JMLR:v11:raykar10a} and Learning from Noisy Labels (LNL)~\citep{10.1023/A:1022873112823,NIPS2013_3871bd64}, following the insights presented by \citet{dawson2021rethinkingnoisylabelmodels}.

Within our framework, we evaluate three state-of-the-art noisy-label learning algorithms: Progressive Label Correction (PLC)~\citep{prog_noise_iclr2021}, DivideMix~\citep{li2020dividemix}, and Self-Evolution Average Label (SEAL)~\citep{chen2021beyond}. 
Based on the results of a preliminary experiments, we select DivideMix for integration into our proposed DINA framework.

DivideMix leverages semi-supervised learning and Gaussian Mixture Models (GMM) to dynamically partition training samples into labeled (clean) and unlabeled (noisy) sets based on their loss distributions. To reduce confirmation bias, DivideMix simultaneously trains two neural networks, each referencing the other's partitions. 
Furthermore, it employs co-refinement and co-guessing techniques inspired by MixMatch to iteratively enhance label accuracy, effectively utilizing both labeled and unlabeled samples to improve model robustness.

DivideMix's effective utilization of unlabeled data, dynamic differentiation between clean and noisy samples, and robust handling of realistic label noise make it particularly suited for countering internal adversarial labeling scenarios. 
In our experiments, we demonstrate that incorporating DivideMix within the DINA framework substantially improves both model robustness and accuracy.

\subsection{Defending Against External Unknown-Word Attacks}
In addition to internal sabotage, external adversarial attacks present another major challenge. 
Unknown-word attacks involve the manipulation of textual input by replacing key words with visually similar but semantically different tokens, making it difficult for conventional LLMs to correctly classify the text. 
To counteract these attacks, we incorporate adversarial training proposed by \citet{yoo-qi-2021-towards-improving}, including the following features: 
\begin{itemize}
    \item \textbf{Generating Adversarial Perturbations}: Introducing adversarially modified examples during training to help the model recognize manipulated patterns.  
    \item \textbf{Robust Word Embeddings}: Training word representations to be more resilient to subtle textual manipulations, allowing the model to correctly interpret adversarial inputs.  
    \item \textbf{Gradient-Based Adversarial Detection}: Using gradient-based methods to identify high-risk modifications in textual input and flag them for review.  
\end{itemize}

By integrating adversarial training into the learning process, we significantly improve the model's resistance to adversarially modified text, ensuring that it correctly detects policy-violating content, even when crafted to evade detection.

\subsection{A Unified Dual Defense Framework (DINA)}
To simultaneously mitigate internal data contamination and external adversarial attacks, we propose a comprehensive hybrid framework named DINA (Dual Defense Against Internal Noise and Adversarial Attacks).
DINA integrates noisy label learning and adversarial training to identify and neutralize adversarially mislabeled data and to enhance model robustness against adversarial text manipulations.

Our four-stage DINA framework extends the three-stage algorithm originally proposed for image recognition by \citet{dawson2021rethinkingnoisylabelmodels}, specifically adapting it to adversarial NLP scenarios.
Our key insight is to proactively counter internal labeling threats through semi-supervised learning and synthesized re-labeling, thus improving intrinsic model robustness. Concurrently, we explicitly strengthen external robustness through adversarial training. The four stages are as follows:

\begin{enumerate}
\item \textbf{Training Weak Learners}:
To prevent initial learners from overfitting to noisy labels, we first train multiple weak learners in the semi-supervised manner on different subsets of the original training corpus.
This stage produces a diverse ensemble of weak learners with better generalization capability, reducing their susceptibility to internal label noise.

\item \textbf{Synthesizing Noisy-Labeled Data}:  
Next, we employ these trained weak learners to generate synthetic noisy labels on additional data samples. 
Using the learning from crowds (LFC) approach~\citep{JMLR:v11:raykar10a}, this synthetic dataset mimics internally poisoned labels realistically and systematically, providing controlled exposure to the type of adversarial noise that the model needs to defend against.

\item \textbf{Main Model Training with Noisy Label Learning}:  
In the third stage, we carefully select samples from the synthesized noisy-labeled dataset generated previously. 
Utilizing DivideMix~\citep{li2020dividemix}, we train the primary NLP moderation model on these selected samples. By doing so, the model learns to robustly differentiate genuine content from internally introduced label noise, effectively suppressing internal adversarial attacks.

\item \textbf{Adversarial Training for External Attack Resilience}:  
Finally, to protect the model against external input-level adversarial perturbations (e.g., synonym substitutions, obfuscations), we conduct adversarial training on the main model with Attacking to Training (A2T)~\citep{yoo-qi-2021-towards-improving}.
Based on BERT-Attack~\citep{li-etal-2020-bert-attack}, we augment the training data with carefully crafted adversarial examples that simulate realistic external attacks. 
This further boosts model robustness, ensuring reliable moderation performance even when encountering manipulated textual inputs.
\end{enumerate}

By integrating these complementary defense strategies, DINA effectively addresses the simultaneous challenges of internal label poisoning and external adversarial perturbations.
As demonstrated in subsequent experimental sections, our approach significantly improves the robustness, accuracy, and generalization capabilities of NLP safety-guard models operating in adversarially complex real-world scenarios.

\section{Experiments}
Our framework is evaluated in real-world corporate scenarios, where we examine whether it effectively protects customer service AI systems from intentional and unintentional sabotage. 
By implementing DINA, we aim to ensure that AI models remain accurate, reliable, and resilient, even in environments where human and AI competition may lead to deliberate interference.  

\subsection{Experimental Setup}

\begin{table}[!t]
    \centering
    \caption{Statistics of our dataset. The training and development sets were curated from real-world data and inherently contain noisy labels provided by discontented annotators. In contrast, the 1,000 test instances were carefully reviewed and reliably annotated by two trusted experts, enabling accurate performance evaluation.}
    \begin{tabular}{lrrr}
    \toprule
    Dataset   &  Spam & Benign & Total \\
    \midrule
    Training Set & 32,127 &  362,554 & 394,681 \\
    Development Set & 3,402 & 46,121 & 49,523\\
    Test Set & 500 & 500 & 1,000 \\
    \bottomrule
    \end{tabular}
    \label{tab:dataset}
\end{table}

We evaluate the effectiveness of our proposed DINA framework using a real-world Chinese dataset curated from an online gaming service company.
Due to annotators' job-security concerns (i.e., the ``discontented annotators''), the provided labels naturally contain significant noise, reflecting realistic internal adversarial scenarios.

As summarized in Table~\ref{tab:dataset}, our dataset consists of a training set (394,681 messages) and a development set (49,523 messages), both inherently noisy due to the original annotators' uncertainty and labeling inconsistencies. 
The spam class includes diverse malicious content, such as advertisements, offensive language, and fraud-related messages.

To accurately evaluate our approach, we further constructed a carefully annotated test set of 1,000 instances, independently reviewed and reliably labeled by two trusted expert annotators. This rigorous annotation procedure eliminates internal label noise, resulting in a balanced evaluation dataset comprising 500 benign and 500 spam instances, serving as a reliable benchmark for assessing model robustness.

We establish a baseline model by fine-tuning the pre-trained \texttt{bert-base-chinese} model~\citep{devlin-etal-2019-bert} for 10 epochs on the noisy training instances. 
To simulate external adversarial scenarios, we apply two attack strategies: (1) Random Attack, which randomly replaces tokens in messages, and (2) the more sophisticated, context-aware BERT-Attack~\citep{li-etal-2020-bert-attack}, which carefully replaces tokens to mislead the model.

\subsection{Results}

\begin{table}[!t]
\centering
\caption{Performance (Accuracy) comparison between the baseline model and our DINA framework under different external adversarial attacks. Results illustrate that DINA, trained to mitigate both internal label noise and external adversarial perturbations, consistently outperforms the baseline model.}
\begin{tabular}{lcc}
\toprule
Type of External Attack & Baseline & DINA \\
\midrule
No External Attack & 0.835 & \textbf{0.903} \\
Random Attack & 0.802 & \textbf{0.901} \\
BERT-Attack~\citep{li-etal-2020-bert-attack} & 0.798 & \textbf{0.862} \\
\bottomrule
\end{tabular}
\label{tab:results}
\end{table}

Experimental results are reported in Table~\ref{tab:results}. 
Without external attacks, our DINA framework achieves an accuracy of 0.903, substantially outperforming the baseline (0.835). This improvement validates the effectiveness of our four-stage training approach in mitigating internal label poisoning.

Under the Random Attack scenario, the baseline performance drops notably from 0.835 to 0.802, highlighting its vulnerability even to simple adversarial token replacements. By contrast, our DINA model demonstrates remarkable robustness, maintaining a high accuracy (0.901) with only negligible degradation.

Finally, the more sophisticated BERT-Attack poses a slightly greater challenge, reducing DINA's accuracy to 0.862. Nevertheless, DINA still significantly surpasses the baseline model's accuracy (0.798). Interestingly, the baseline model suffers similarly under both random and context-aware attacks, indicating its inherent susceptibility even to naïve adversarial manipulations. Our results clearly demonstrate the dual robustness of DINA, effectively defending against both internal label noise and external adversarial attacks.

\subsection{Impact of Internal Label Noise on Model Robustness}
\begin{figure}[!t]
    \centering
    \includegraphics[width=1\linewidth]{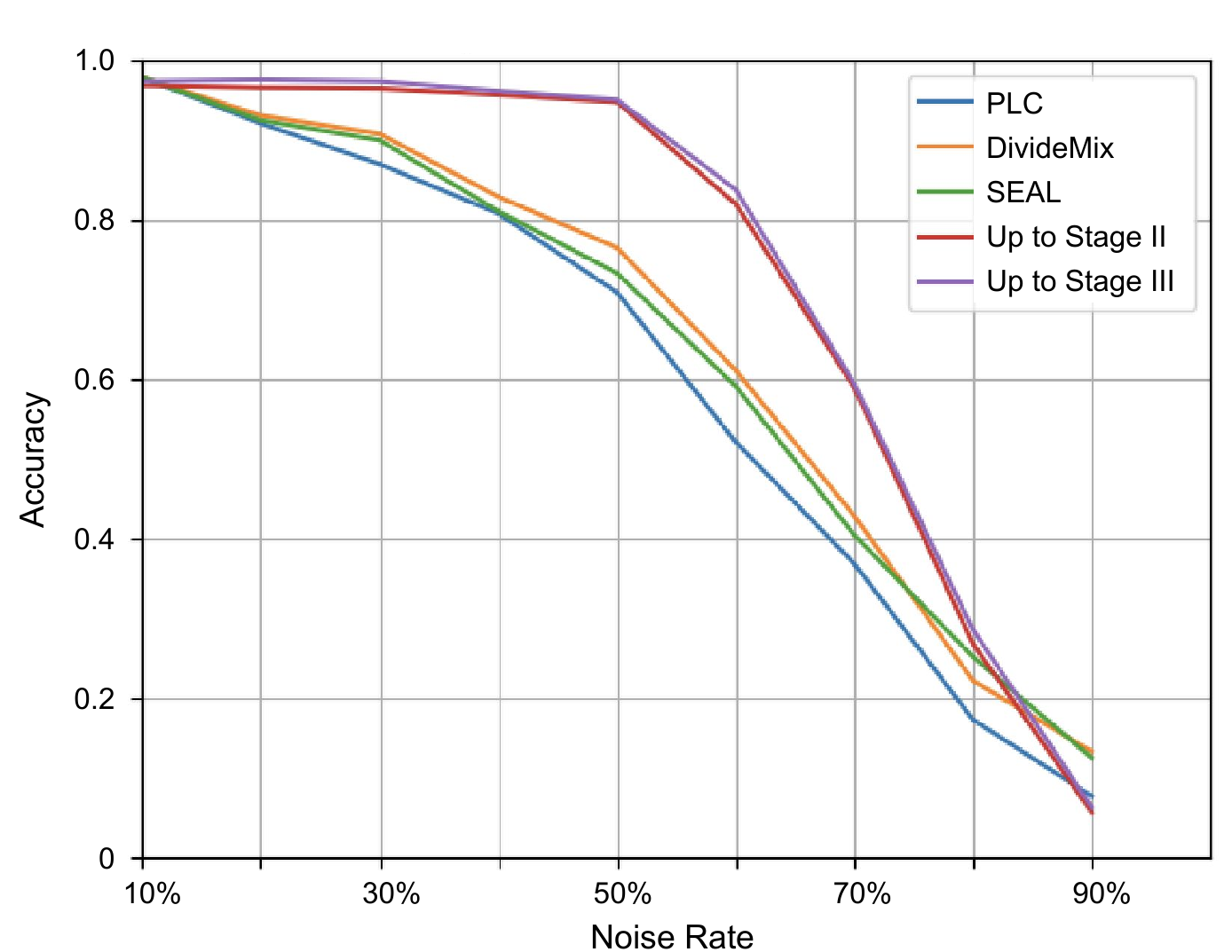}
    \caption{Performance comparison (Accuracy) of different noisy-label learning (LNL) methods under varying levels of internal label noise. The graph compares Progressive Label Correction (PLC), DivideMix, Self-Evolution Average Label (SEAL), and our DINA framework at Stage II (with LFC only) and Stage III (LFC + DivideMix). 
    Results indicate that DivideMix outperforms other existing LNL methods, while integrating LFC (Stage II) significantly enhances robustness. Further incorporating DivideMix (Stage III) achieves additional incremental improvements.}
    \label{fig:lnl-analysis}
    \Description{Performance comparison (Accuracy) of different noisy-label learning (LNL) methods under varying levels of internal label noise. The graph compares Progressive Label Correction (PLC), DivideMix, Self-Evolution Average Label (SEAL), and our DINA framework at Stage II (with LFC only) and Stage III (LFC + DivideMix). 
    Results indicate that DivideMix outperforms other existing LNL methods, while integrating LFC (Stage II) significantly enhances robustness. Further incorporating DivideMix (Stage III) achieves additional incremental improvements.}
\end{figure}

To assess the robustness of our model under different intensities of internal label sabotage, we simulated various noise levels by flipping labels at predefined ratios (ranging from 10\% to 90\%) in the training set. 
We then evaluated the performance of multiple noisy-label learning (LNL) methods on the development set. Figure~\ref{fig:lnl-analysis} presents the accuracy comparisons among Progressive Label Correction (PLC)\citep{prog_noise_iclr2021}, DivideMix\citep{li2020dividemix}, Self-Evolution Average Label (SEAL)~\citep{chen2021beyond}, and two intermediate versions of our DINA framework: Stage-II (with only Learning from Crowds, LFC) and Stage-III (LFC combined with DivideMix-based LNL).

The results clearly demonstrate that DivideMix consistently achieves superior performance compared to other individual LNL methods when the noise rate under 70\%. 
Our framework, even at Stage-II (LFC alone), substantially outperforms these existing methods across most noise rates, underscoring the critical role of crowdsourced relabeling. 
Further integrating DivideMix at Stage-III yields incremental yet meaningful improvements, indicating the complementary advantage of combining crowdsourced relabeling with advanced noisy-label learning to mitigate internal label noise effectively.

\subsection{Impact of External Adversarial Perturbations on Model Robustness}
\begin{figure}[!t]
    \centering
    \includegraphics[width=1\linewidth]{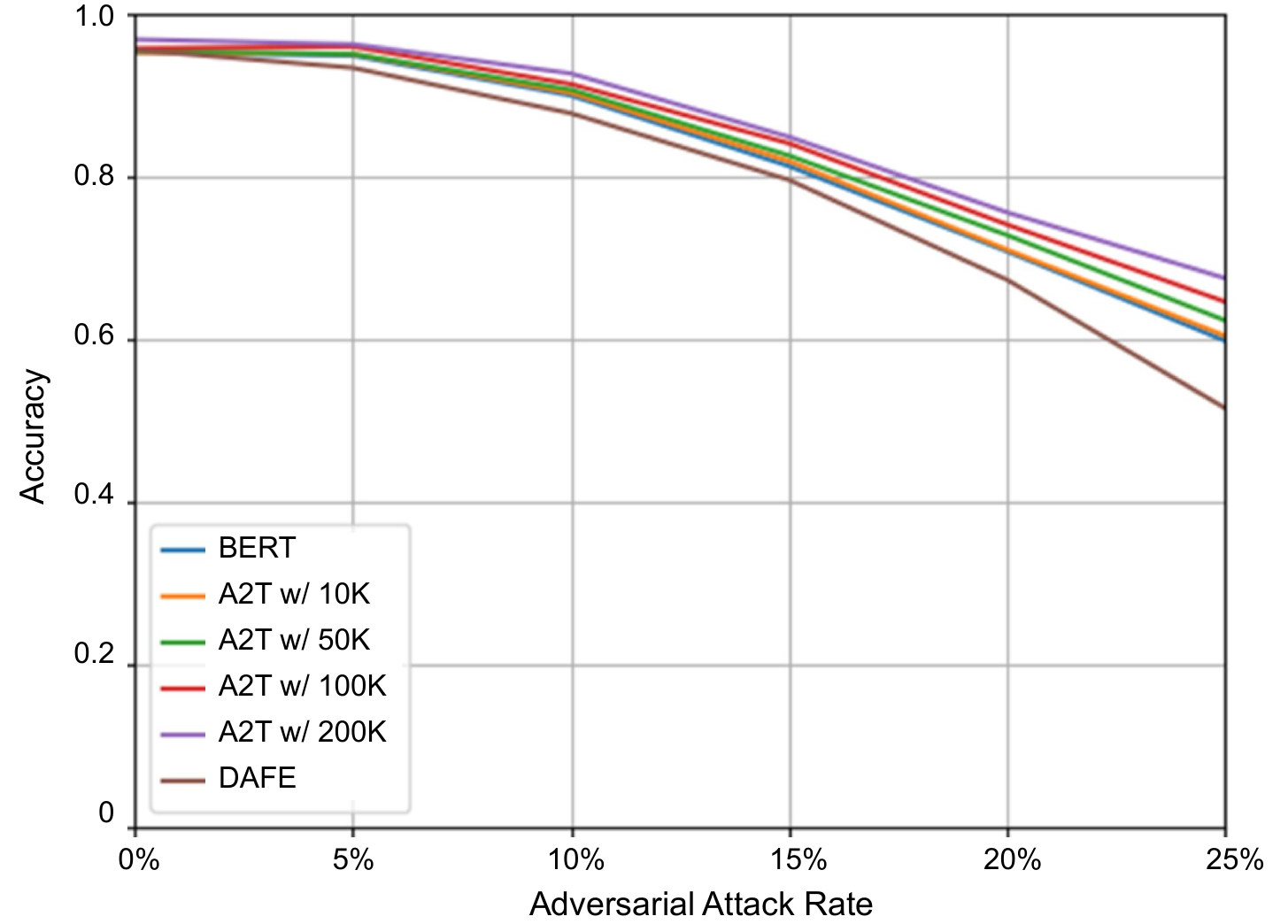}
    \caption{Analysis of the impact of adversarial example quantity on model performance under the Attacking-to-Training (A2T) strategy~\citep{yoo-qi-2021-towards-improving}. 
    We report accuracy for our DINA framework with different quantities of adversarial examples, compared to the baseline BERT model and the unsupervised domain-transfer approach DAFE~\citep{dou-etal-2019-unsupervised}. A2T achieves optimal performance at approximately 200K adversarial examples, demonstrating its effectiveness in enhancing model robustness.}
    \label{fig:a2t-analysis}
    \Description{Analysis of the impact of adversarial example quantity on model performance under the Attacking-to-Training (A2T) strategy. Accuracy is reported for our DINA framework with varying adversarial example sizes, compared to the baseline BERT model and the unsupervised domain-transfer approach DAFE~\citep{dou-etal-2019-unsupervised}. A2T achieves optimal performance at approximately 200K adversarial examples, demonstrating its effectiveness in enhancing model robustness.}
\end{figure}

We further analyze the impact of the number of adversarial examples utilized in the Attacking-to-Training (A2T) adversarial training strategy employed during Stage IV of our DINA framework. Figure~\ref{fig:a2t-analysis} compares the performance of the A2T strategy with varying sizes of adversarial examples against both the baseline BERT model and an unsupervised domain-adaptation approach, domain-aware feature embeddings (DAFE)~\citep{dou-etal-2019-unsupervised}.
For DAFE, we treat perturbed adversarial messages as the target domain and perform unsupervised domain transfer to adapt the model accordingly.

Our results clearly indicate that A2T achieves optimal performance when trained on approximately 200K adversarial examples. 
In contrast, DAFE consistently exhibits the lowest accuracy, underscoring its unsuitability for mitigating external adversarial perturbations in our scenario. 
This analysis highlights the effectiveness of the A2T adversarial training strategy within our DINA framework, particularly in bolstering the model's robustness against realistic adversarial text manipulations.

Through extensive experiments, we demonstrate that DINA successfully mitigates internal and external threats, maintaining high performance despite adversarial disruptions. Our research provides a practical and scalable defense strategy for real-world NLP applications, offering valuable insights into the challenges of AI-human competition and adversarial robustness.

\section{Conclusion}
In this paper, we introduced DINA (Dual Defense Against Internal Noise and Adversarial Attacks), a unified framework designed to simultaneously counteract internal adversarial labeling and external adversarial text perturbations in NLP applications. By integrating semi-supervised learning, crowdsourced relabeling, advanced noisy-label learning, and adversarial training, DINA effectively addresses the dual threats posed by human-driven adversarial behaviors.

Experimental results on real-world data from an online gaming service demonstrate that DINA significantly outperforms baseline models, effectively mitigating both internal label noise and external adversarial attacks. Specifically, our framework achieved optimal robustness when trained with approximately 200K adversarial examples, highlighting the practical applicability of our approach.

Overall, this study underscores the critical importance of explicitly addressing simultaneous adversarial threats in NLP systems. By providing a comprehensive and novel integration of noisy label learning and adversarial training, our approach not only enhances model robustness, accuracy, and stability under realistic adversarial conditions but also offers practical solutions to societal tensions between human workers and AI systems. Future work includes extending our dual-defense framework to additional NLP tasks and exploring broader classes of adversarial perturbations.

\begin{acks}
This research was partially supported by the National Science and Technology Council (NSTC), Taiwan, under Grant Nos. 112-2221-E-001-016-MY3 and 113-2926-I-004-001-MY3, and by Academia Sinica under Grant No. 236d-1120205.
\end{acks}

\bibliographystyle{ACM-Reference-Format}
\bibliography{dina}


\end{document}